\documentclass[fleqn,12pt,twoside]{article}
\usepackage{espcrc1_qm02}

\usepackage{graphicx}

\usepackage{epsfig}

\usepackage{amssymb}

\newcommand{\AmS}{{\protect\the\textfont2
  A\kern-.1667em\lower.5ex\hbox{M}\kern-.125emS}}

\def\psat{p_{\rm sat}}

\def\ptav{$\langle p_T\rangle$}

\def\Teff{T_{\rm eff}}

\def\Tdec{T_{\rm dec}}

\newcommand{\beq}{\begin{equation}}
\newcommand{\eeq}{\end{equation}}
\newcommand{\bea}{\begin{eqnarray}}
\newcommand{\eea}{\end{eqnarray}}

\long\def\comment#1{ }        

\title{Transverse spectra of hadrons at RHIC}

\author{K.~J.~Eskola\address[JYFL]
        {Department of Physics,  University of Jyv\"askyl\"a,\\
        P.O. Box 35, FIN-40014 University of Jyv\"askyl\"a, Finland},
        H.~Niemi\addressmark[JYFL],
        \underline{P.~V.~Ruuskanen}\addressmark[JYFL],
        S.~S.~R\"as\"anen\addressmark[JYFL]}

\begin{document}

\maketitle

\begin{abstract}
We present results on spectra of pions, kaons and (anti)protons from a
study of heavy ion collisions using the perturbative QCD + saturation
model to calculate the production of initial (transverse) energy and
baryon number followed by a hydrodynamic description of the expansion
of produced matter.  In particular, we study how the hadron spectra
and multiplicities depend on the decoupling temperature $\Tdec$ when
the low temperature phase contains all hadrons and and hadron
resonances with mass below 2 GeV.  We show that the spectra and
multiplicities of pions, kaons and (anti)protons measured at RHIC in
central Au+Au collisions with $\sqrt s=130$ GeV can be obtained with a
single decoupling temperature 150$\dots$160 MeV, common for both the
chemical and the kinetic freeze-out.
\end{abstract}


\section{INTRODUCTION}
\label{Intro}

The average transverse momentum of primary particles in heavy ion
collisions is expected to be larger than the experimentally observed
one.  Two ways out of this problem have been discussed: In one case
the number of produced particles is small
and the final multiplicity is achieved after fragmentation which
reduces also the average transverse momenta of final, observed
particles \cite{MUELLER}.  On the other hand, results from the
calculation can be used as initial conditions in a hydrodynamical
approach \cite{EKRT}.  In this case the extra transverse energy is
transferred into the longitudinal motion through the work by pressure
in the expansion.  The transfer from transverse to longitudinal
degrees of freedom takes place because the matter is produced in a
strong longitudinal expansion.

When time passes, also a transverse flow builds up and leads, e.g., to a
clear mass dependence of final hadron spectra.  The shape of spectra
depend both on the transverse flow and the decoupling temperature
$\Tdec$ but the ratios of particle multiplicities on the $\Tdec$ and on
the baryon chemical potential $\mu_{B,{\rm dec}}$.  Strangeness chemical
potential is fixed by demanding that the net strangeness density is
zero.

Modelling the heavy ion collisions at collider energies in terms of
perturbative QCD calculation of minijet production at saturation
\cite{EKRT} and of hydrodynamic expansion of produced matter, 
has lead to a successful description of 
multiplicities and transverse energy at
mid-rapidities \cite{ERRT}.  This talk summarizes the predictions for
particle spectra in central Au+Au collisions at $\sqrt s=130$~GeV from
the pQCD+saturation+hydrodynamics model \cite{ENRR}.

\section{THE MODEL FOR CALCULATING THE TRANSVERSE SPECTRA}
\label{Model}
In a high energy $AA$ collision parton production can be calculated from
the nuclear parton distribution functions and the perturbative parton-parton
cross sections, provided that the momentum transfer in the partonic
collision $p_T\gg\Lambda_{\rm QCD}$.  Assuming independent partonic
collisions, the dominant part of the initial production is obtained by
extending the (LO) calculation of partonic jets down to
$p_0=1\dots2$~GeV$\gg\Lambda_{\rm QCD}$ \cite{EKL}.

The basic quantities for characterizing the parton production in nuclear
collisions are the number and the total transverse energy of minijets
per unit rapidity.  These are obtained as a function of $p_0$ from the
transverse energy distribution of minijets in the central rapidity unit
as the normalization $\sigma_{\rm jet}(p_0)$ and the first moment
$\sigma\langle E_T\rangle$ \cite{EKL}.  Combining these with the nuclear
overlap function $T_{AA}$ (or the product of thickness functions of
colliding nuclei) gives the average total number and transverse energy
of partons with $p_{\rm T}>p_0$.  The net baryon number can be obtained
as a difference of the quark and anti-quark numbers \cite{EK}.  Finally,
to close the model, a cut-off momentum must be fixed and here we assume
that it originates from the saturation of partons \cite{GLR}.  We
express this in terms of final, produced partons as a simple geometrical
condition:  each parton is given a transverse size $\pi/p_T^2$, and when
the partons overlap, extra independent primary collisions with smaller
$p_T$ are suppressed.  The lower cut-off $p_0=\psat$ is thus fixed from
the saturation condition \cite{EKRT} $N(p_0,\Delta y)\cdot \pi/p_0^2=\pi
R_A^2$.

{From} the minijets at saturation we can calculate the energy density
and the net baryon number density after introducing a correlation
between the longitudinal momentum and the longitudinal space-time
formation point of the minijet.  We assume that rapidity of the
minijet equals the space-time rapidity of the formation point,
$y=\eta=0.5\ln[(t+z)/(t-z)]$.  The formation time, defined as $1/\psat$
is taken as the initial time for the hydrodynamic evolution.  The
details of the determination of the initial state can be found in
\cite{ENRR,ERRT}.

It can be argued that $\tau_0=1/\psat$ ($\sim$0.2 fm/$c$ for Au+Au at
RHIC) is too short to achieve thermalization. However, from the point
of view of the average energy per particle, the system does look
thermal \cite{EKRT}.  Also, one should expect collisions to take place
and the build-up of collective behaviour to start already during the
thermalization process.  From this point of view, the use of
hydrodynamics at early times should be considered to approximate the
effects of momentum transfer in the collisions during equilibration.

To solve the hydrodynamic equations, an equation of state (EoS) is
needed.  We assume an ideal QGP high temperature phase with a first
order phase transition to a hadron resonance gas at $T_c=165$ MeV. All
hadrons and hadron resonances with $M<2$ GeV are included and a
repulsive mean field is assumed in order to ensure a consistent
temperature behaviour of the pressure \cite{Sollfrank}. The final
spectra are obtained using the Cooper--Frye freeze-out procedure and
the two- and three-body decays of all resonances are included.

\section{RESULTS}
\label{Res}

In Figure 1 the $\Tdec$ dependence of spectra of negative pions, kaons
and antiprotons is shown together with the data measured by the PHENIX
Collaboration \cite{PHENIX}.  The slopes of the spectra are seen to
change considerably with $\Tdec$.  For $p_T>1.5$ GeV, the decrease of
$\Tdec$ from 150 MeV to 120 MeV changes the inverse slope $\Teff$ from
265 MeV to 305 MeV for pions, from 275 MeV to 330 MeV for kaons and
from 295 MeV to 375 MeV for protons.  For the \ptav\ the changes are
from 0.46, 0.65 and 0.86 GeV to 0.53, 0.78 and 1.08 GeV for pions,
kaons and protons, respectively.  The change of the decoupling
temperature affects also the normalization of spectra, that is the
multiplicity.  In fact the multiplicity of pions increases slightly
whereas that of kaons decreases 21~\% and of protons 52~\%.  As has
been observed earlier \cite{REDLICH}, to reproduce in thermal models
the multiplicities of heavier particles and pions simultaneously, a
freeze-out temperature for chemical reactions must be of the order of
150 to 160 MeV. In our calculation not only the multiplicities but
also the observed shapes of the spectra are obtained if the kinetic
freeze-out and the chemical freeze-out are assumed to take place at
the (approximately) same temperature of order 150$\dots$160 MeV.

 \vspace{-1.cm}

\begin{figure}[h!bt]
\vspace{0.3cm}
\begin{minipage}[t]{75mm}
  \begin{center}
    \epsfxsize 75mm \epsfbox{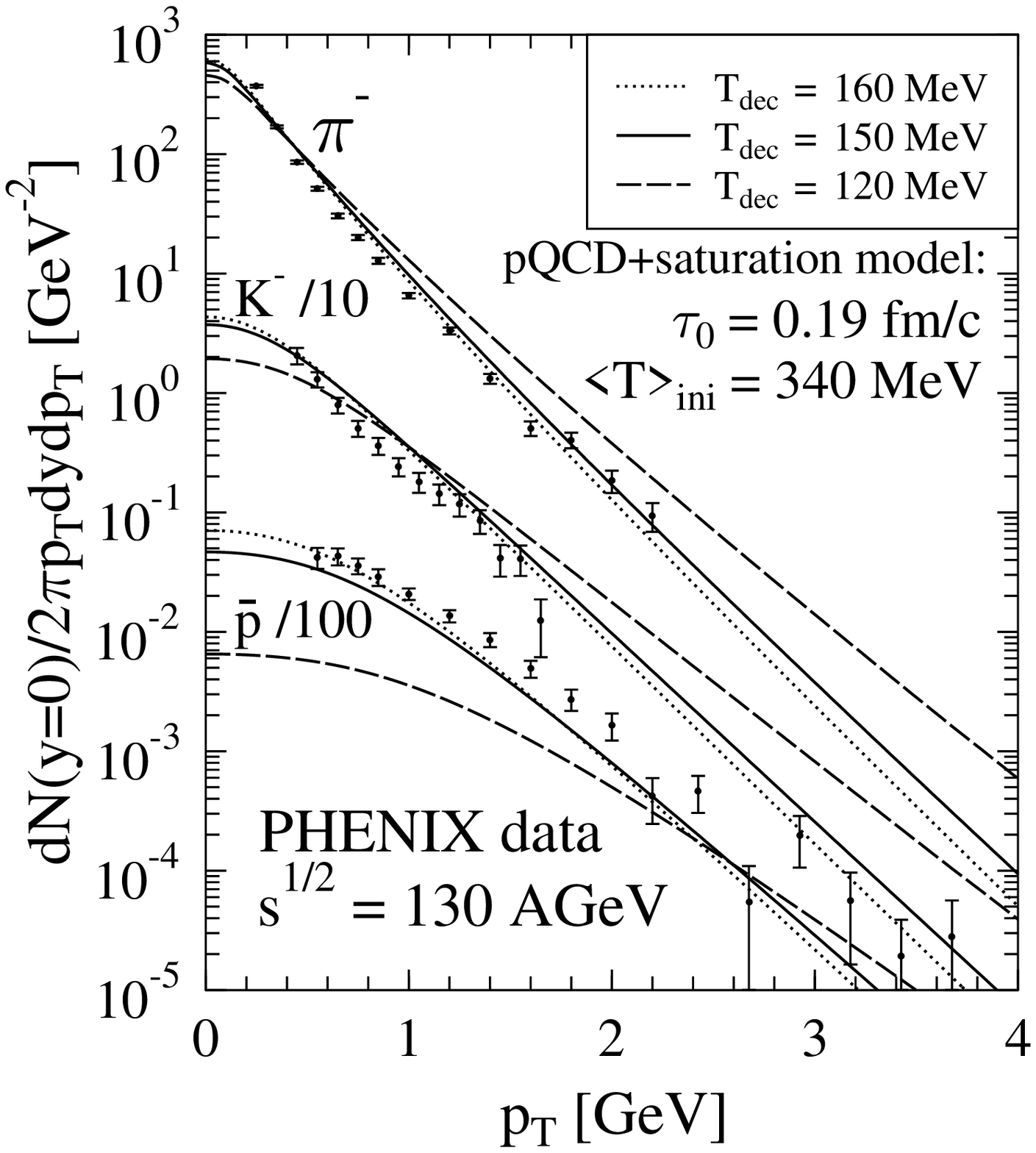}
  \end{center}
 \vspace{-1.7cm}
 \caption{\small Spectra of negative particles calculated \cite{ENRR} 
	  for $\Tdec=$ 160, 150 and 120 MeV. 
          The data points are from a measurement by the PHENIX collaboration
          \cite{PHENIX}.
          }
 \label{negative}
\end{minipage}
\hspace{\fill}
\begin{minipage}[t]{75mm}
  \begin{center}
    \epsfxsize 75mm  \epsfbox{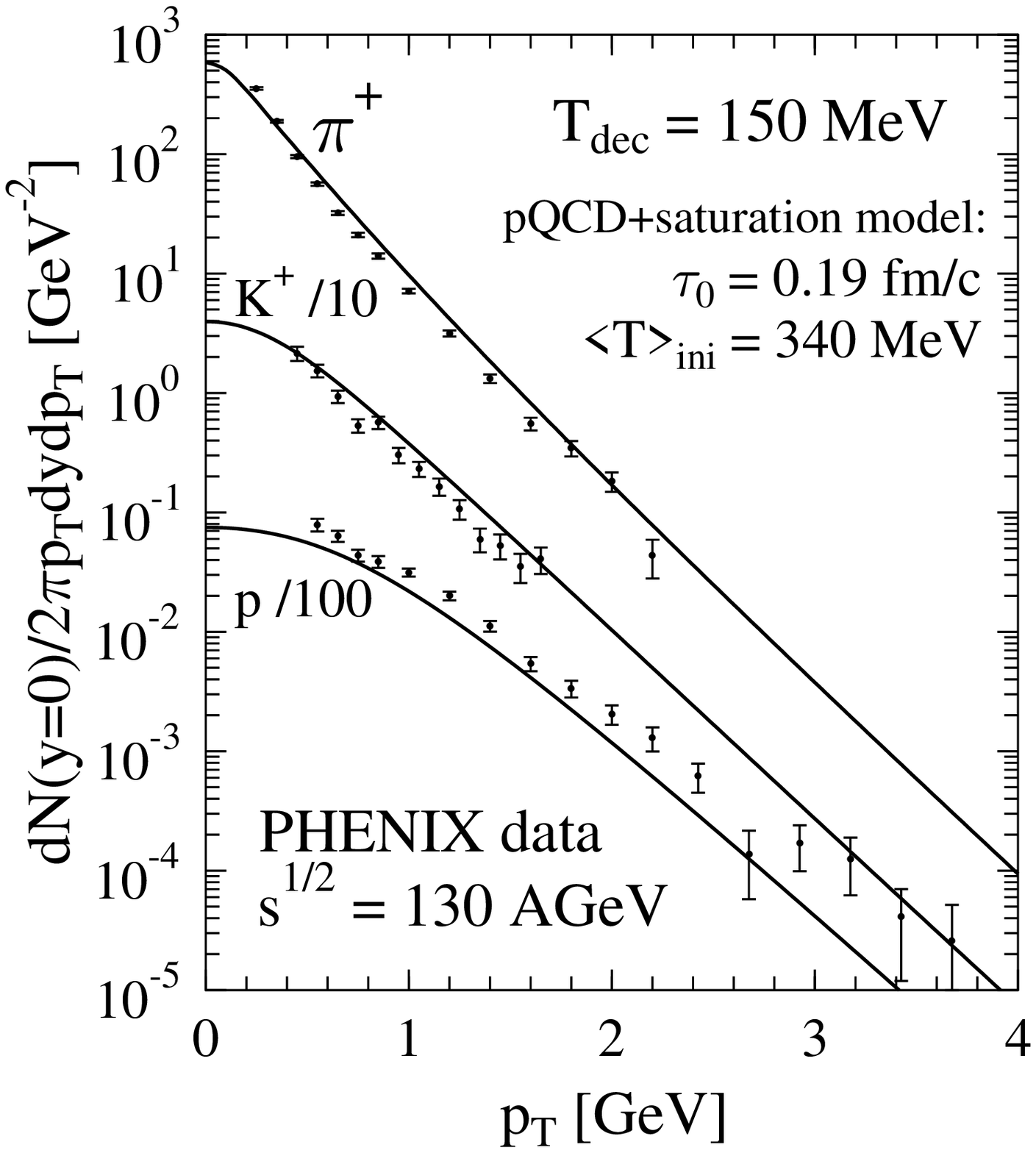}
  \end{center}
 \vspace{-1.7cm}
 \caption{\small As in Fig.~1 but for positive particles and with
          $\Tdec = 150$ MeV only.
          The feed-down from hyperon states is
          included in the data but not in the calculation. }
 \label{positive}
\end{minipage}
\vspace{-0.4cm}
\end{figure}

In Figure 2 the spectra of positive pions, kaons and protons are
depicted using $\Tdec=150$ MeV in the calculation.  Within the accuracy
of the data the normalizations of pions and kaons agree but the
calculated proton spectrum is slightly below the measured one.  This is
to be expected since the feed-down from hyperons is included among the
measured but not the calculated protons.  PHENIX has also reported $p$
and $\bar p$ yields corrected for hyperon feed-down, $dN/dy=19.3\pm0.6$
and $13.7\pm0.7$ \cite{PHENIX_lambda}.  Our results for $\Tdec=150$ MeV
are 20.2 and 13.1 in excellent agreement with the measurement.  Also the
ratio $\bar\Lambda/\Lambda=0.70$ is consistent with the PHENIX result
$0.75\pm0.09$ \cite{PHENIX_lambda}.

Note that the deviation of $\bar p/p$ and $\bar\Lambda/\Lambda$ from
unity is due to the net baryon number content of the initial matter at
saturation. In the pQCD + saturation calculation the net baryon number
cannot be changed independently of the total multiplicity.  The
relative production of quarks and antiquarks (and gluons) is
completely fixed by the parton distributions and the perturbative
parton level cross sections. In addition to the net baryon number, the
number of baryons relative to that of antibaryons depends strongly on
decoupling temperature and again in our calculation agreement is
obtained only for $\Tdec\simeq$~150$\dots$160~MeV.

\section{DISCUSSION} \label{Disc}
Using the pQCD + saturation model \cite{EKRT} to calculate the initial
particle production in central Au+Au collisions at $\sqrt s = 130$ GeV
and hydrodynamics to describe the expansion of the matter we obtain a
good agreement between the calculated and measured quantities.  We
find it remarkable that in order to reproduce either the
multiplicities, the slopes or the antibaryon-to-baryon ratios we come
up in each case with the same decoupling temperature of order
150$\dots$160 MeV.  Similar conclusion was also obtained in
\cite{Florkowski}.

It has been argued that $\tau_0=1/\psat \sim 0.2$ fm/c is too short a
time for achieving full thermalization. This might well be so but it
is not plausible to assume that there are no collisions with momentum
transfers among the constituents at times before a possible later
thermalization time. Collective motion, the flow, is a result of the
collisions among the particles of the matter which lead to a net
momentum transfer from denser to less dense parts in the matter. From
this point of view it is difficult to justify the starting of a
hydrodynamical calculation at a later time assuming no initial
transversal velocity. In any case, we have checked the effect of the
initial time on the shapes of the spectra.  Taking $\tau_0=0.6$ fm/c,
needed to explain the formation of elliptic flow \cite{Kolb},
while keeping the total entropy fixed,  changes the slopes only
slightly. Even $\tau_0=1.0$ fm/c is not enough to compensate for the
change from the decrease of $\Tdec$ from 150 to 120 MeV.

{Finally}, we want to emphasize that our model scheme is very tight.
Uncertainties in the pQCD + saturation calculation are essentially
fixed from a single global quantity, like the total multiplicity.  In
the hydrodynamical part a large number of hadrons and hadron
resonances must be included but after fixing the EoS, the only real
freedom left is the decoupling temperature $\Tdec$.


\vspace{0.2cm}\noindent
{\bf Acknowledgement.} {\small We thank the Academy of Finland, Project 50338,
for financial support.}

\end{document}